\documentclass[a4paper]{article}

\usepackage{INTERSPEECH2020}
\usepackage{epstopdf,amsmath}
\usepackage{epsfig}

\title{Neural Discriminant Analysis for Deep Speaker Embedding}
\name{Lantian Li, Dong Wang, Thomas Fang Zheng}
\address{
  Center for Speech and Language Technologies, Tsinghua University, China}
\email{lilt@cslt.org; wangdong99@mails.tsinghua.edu.cn; fzheng@tsinghua.edu.cn}

\begin{document}

\maketitle
\begin{abstract}

Probabilistic Linear Discriminant Analysis (PLDA) is a
popular tool in open-set classification/verification tasks. However, the Gaussian assumption underlying PLDA
prevents it from being applied to situations where the data is clearly non-Gaussian.
In this paper, we present a novel nonlinear version of PLDA named as Neural Discriminant Analysis (NDA).
This model employs an invertible deep neural network to transform a complex distribution to a simple Gaussian,
so that the linear Gaussian model can be readily established in the transformed space.
We tested this NDA model on a speaker recognition task where the deep speaker vectors (x-vectors) are
presumably non-Gaussian. Experimental results on two datasets demonstrate that NDA consistently outperforms PLDA,
by handling the non-Gaussian distributions of the x-vectors.


\end{abstract}

\section{Introduction}

Probabilistic Linear Discriminant Analysis (PLDA)~\cite{Ioffe06,prince2007probabilistic}
has been used in a wide variety of recognition tasks,
such as speaker recognition (SRE)~\cite{kenny2010bayesian}.
In nearly all the situations, PLDA cooperates with a \emph{speaker embedding} front-end, and
plays the role of scoring the similarity between one or a few enrollment utterances and the test utterance,
represented in the form of speaker vectors. Traditional speaker embedding approaches are based on statistical models,
in particular the i-vector model~\cite{dehak2011front}, and recent embedding approaches are based on
deep neural networks (DNNs)~\cite{ehsan14,li2017deep}, for
which the x-vector model~\cite{snyder2018xvector,okabe2018attentive,cai2018exploring} is the most successful.

PLDA is a full-generative model, based on two primary assumptions:
(1) all the classes are Gaussians and these Gaussians share the same covariance matrix;
(2) the class means are distributed following a Gaussian.
The full-generative model offers a principle way to deal with classification tasks where
the classes are represented by limited data. Taking SRE as an example, in most cases,
a speaker registers itself with only one or a few enrollment utterances, which means that the distribution of the
speaker is
not fully represented and the identification/verification has to be conducted base on an
uncertain model. The PLDA model solves this problem in a Bayesian way:
it represents each speaker as a Gaussian with an \emph{uncertain} mean, and computes the
likelihood of a test utterance by marginalizing the uncertainty.
Due to this elegant uncertainty treatment, PLDA has been widely used in SRE,
and has achieved state-of-the-art performance in many benchmarks, in particular when combined
with length normalization~\cite{garcia2011analysis}.


Although promising, PLDA suffers from a limited representation capability.
Specifically, PLDA is a linear Gaussian model, and the prior, the conditional, and the marginal are all assumed to be Gaussian.
This means that if the distribution of speaker vectors do not satisfy this assumption,
PLDA cannot represent the data well, leading to inferior performance.
This problem is not very serious for speaker vectors that are derived from statistical models where the speaker vectors have
been assumed to be Gaussian, e.g., the JFA model~\cite{Kenny07,kenny2005joint} and the i-vector model~\cite{dehak2011front}.
However, for speaker vectors derived from models that do not possess such a Gaussian assumption, PLDA may be biased.
This is particularly the case for x-vectors~\cite{snyder2018xvector}: they are derived from a deep neural network and
there is not any explicit or implicit Gaussian constraint on the prior and the conditional.

In fact, the importance of Gaussianality of the data for PLDA has been noticed for a long time.
For example, Kenny et al.~\cite{kenny2010bayesian} found that i-vectors exhibit a heavy-tail property and
so are not appropriate to be modeled by Gaussians. They presented a heavy-tail PLDA where the prior and the conditional
are set to be Student's t-distributions. Garcia et al~\cite{garcia2011analysis} found a simple length normalization can improve
Gaussianality of i-vectors, and the traditional PLDA model can recover the performance of the heavy-tail PLDA if the
test i-vectors are length-normalized.

The non-Gaussianality of x-vectors were recognized by Li et al.~\cite{li2019gaussian}.
They found that x-vectors exhibit more complex within-class
distributions compared to i-vectors, and confirmed the non-Gaussianality of x-vectors by computing
the Skewness and Kurtosis of the within-class and between-class distributions. Further analysis
was conducted in~\cite{zhang2019vae}, and a VAE model was used to perform normalization for x-vectors.
Recently, a more complex normalization technique based on a normalization flow was conducted
by Cai et. al~\cite{cai2020deep}. All these studies try to produce more Gaussian speaker vectors,
so that PLDA model can be employed more effectively.

\begin{figure}[htbp]
\centering
\includegraphics[width=1.0\linewidth]{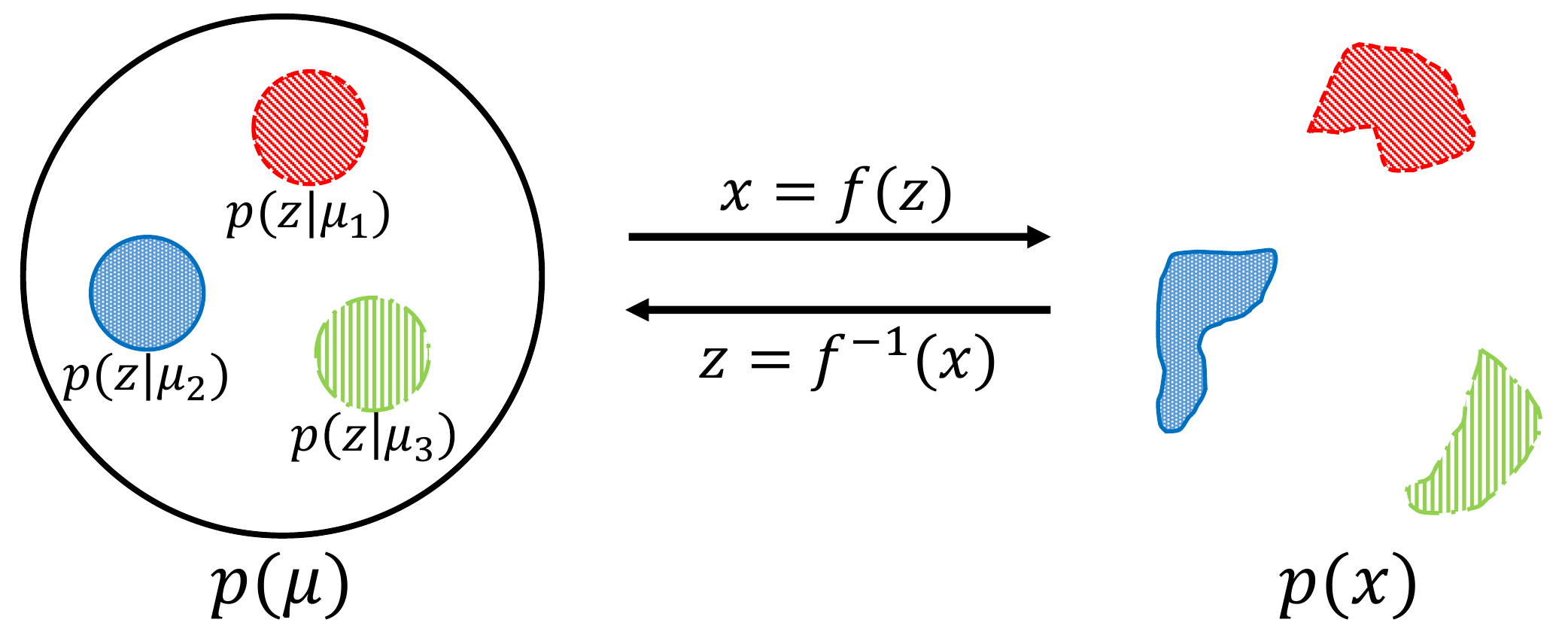}
\caption{Neural discriminant analysis. The complex within-class distributions (on the right) are transformed to Gaussians
with shared covariance by an invertible transform, and then the linear Bayesian model is established in the transformed space.
Each color represents an individual class (speaker).}
\label{fig:nda}
\end{figure}

All the approaches mentioned above cannot be regarded as perfect. The heavy-tailed PLDA and the length normalization
cannot deal with complex distributions, and the normalization techniques do not optimize the
normalization model and the scoring model in a joint way. In this paper, we present a
Neural Discriminant Analysis (NDA) model that can deal with non-Gaussian data in a principle way.
It keeps the strength of PLDA in handling uncertain class representations, while offers the capability to
represent data with complex within-class distributions. The central idea, as shown in Figure~\ref{fig:nda},
comes from the distribution transformation theorem~\cite{rudin2006real}, by which any form of non-Gaussian
distribution can be transformed to a Gaussian, and so a linear Bayesian model can be established
on the transformed variables. We tested the NDA model on SRE tasks with two datasets, VoxCeleb and CNCeleb,
and achieved rather promising results.

The organization of this paper is as follows.
Section~\ref{sec:method} presents the NDA model, and experiments are reported in Section~\ref{sec:exp}.
The paper is concluded in Section~\ref{sec:con}.

\section{Method}
\label{sec:method}

\subsection{Revisit PLDA}

PLDA models the generation process of multiple classes. For each class, the generation process is factorized into two steps:
firstly sample a class mean $\pmb{\mu}$ from a Gaussian prior, and secondly sample a class
member $\pmb{x}$ from a conditional which is another Gaussian centered on $\pmb{\mu}$.
Without loss of generality, we shall assume the prior is diagonal and the conditional is
isotropic\footnote{It can be easily verified that a linear transform does not change the
PLDA score, which makes it possible to transform any form of Gaussians of the prior and the conditional
to the form of Eq.~(\ref{eq:pmu}) and Eq.~(\ref{eq:px-mu}).
Moreover, PLDA with a low-rank loading matrix can be regarded as a special
form of Eq.~(\ref{eq:pmu}), by setting some of the diagonal elements of $\pmb{\epsilon}$ to be 0.}.
This leads to the following probabilistic model:

\begin{equation}
\label{eq:pmu}
p({\pmb{\mu}}) = N({\pmb{\mu}}; \mathbf{0},  \pmb{\epsilon} \mathbf{I})
\end{equation}

\begin{equation}
\label{eq:px-mu}
p({\pmb{x}}|{\pmb{\mu}}) = N({\pmb{x}}; {\pmb{\mu}}, \sigma \mathbf{I}),
\end{equation}

\noindent where $\pmb{\epsilon}\mathbf{I}$ and $\sigma\mathbf{I}$ are the between-class and within-class covariances, respectively.

With this model, the probability that $\pmb{x}_1,...,\pmb{x}_n$ belong to the same speaker can be computed by:

\begin{equation}
\label{eq:pxxx}
  \begin{split}
    p(\pmb{x}_1, ..., \pmb{x}_n) &= \int p(\pmb{x}_1, ..., \pmb{x}_n|\pmb{\mu}) p(\pmb{\mu}) \rm{d} \pmb{\mu}\\
     \propto & \prod_j \epsilon_j^{-1/2} \prod_j (\frac{n}{\sigma} + \frac{1}{{\epsilon_j}})^{-1/2} \\
    & \exp \Big\{ - \frac{1}{2\sigma} \big\{ \sum_i ||\pmb{x}_i||^2 - \sum_j \frac{n^2 \epsilon_j}{n \epsilon_j + \sigma} \widehat{{x}}_j^2 \big\} \Big\},
  \end{split}
\end{equation}

\noindent where $j$ indexes the dimension.

For speaker recognition, our essential goal is to estimate the probability $p(\pmb{x}|\pmb{x}_1, ..., \pmb{x}_n)$, where $\pmb{x}$ is the
test data and $\pmb{x}_1, ..., \pmb{x}_n$ represent the enrollment data. Focusing on the verification task, the above probability should
be normalized by a background probability that $\pmb{x}$ belongs to any possible classes. This leads to a normalized likelihood (NL) for speaker
verification:

\[
R(\pmb{x}; \pmb{x}_1, ..., \pmb{x}_n) = \frac{p(\pmb{x}|\pmb{x}_1, ..., \pmb{x}_n)}{p(\pmb{x})}.
\]

It is easy to verify that the above NL score is equal to the likelihood ratio (LR), the standard form for PLDA scoring in the seminal papers~\cite{Ioffe06,prince2007probabilistic}:

\[
R(\pmb{x}; \pmb{x}_1, ..., \pmb{x}_n) = \frac{p(\pmb{x},\pmb{x}_1, ..., \pmb{x}_n)}{p(\pmb{x}) p(\pmb{x}_1, ..., \pmb{x}_n)},
\]

\noindent and all the terms in the numerator and denominator can be computed by Eq.~(\ref{eq:pxxx}).

\subsection{Neural discriminant analysis (NDA)}

All the derivations for the PLDA model are based on the Gaussian assumption on the prior
and conditional distributions, shown by Eq.~(\ref{eq:pmu}) and Eq.~(\ref{eq:px-mu}). We shall
relax this assumption by introducing a new probabilistic discriminant model based on neural net and Bayesian inference.

\subsubsection{Distribution transform}

Suppose an invertible transform ${f}$ maps a variable $\pmb{z}$ to $\pmb{x}$, i.e., $\pmb{x} = {f}(\pmb{z})$.
According to the principle of distribution transformation for continuous variables~\cite{rudin2006real}, we have:

\begin{equation}
\label{eq:nf}
p({\pmb{x}}) = p({\pmb{z}}) |\frac{\partial{f}^{-1}({\pmb{x}})}{\partial{{\pmb{x}}}}|,
\end{equation}

\noindent where the second term is the absolute value of the determinant of the Jacobian matrix of ${f^{-1}}$, the inverse transform of ${f}$.
This term reflects the change of the volume of the distribution with the transform, and is often called the \emph{entropy term}.
For simplicity, we will denote this term by $J_{{\pmb{x}}}$:

\begin{equation}
\label{eq:nf-2}
p(\pmb{x}) = p(\pmb{z}) J_{\pmb{x}}.
\end{equation}

\noindent Note that $p(\pmb{x})$ and $p(\pmb{z})$ are in different distribution forms.
If we assume $p(\pmb{z})$ to be a Gaussian, then this (inverse) transform plays a role of Gaussianization~\cite{chen2001}.
It was demonstrated that if $f$ is complex enough, any complex $p(\pmb{x})$ can be transformed to a Gaussian or a uniform distribution~\cite{papamakarios2019normalizing}.
Recently, there is numerous research on designing more efficient transform functions by deep neural networks,
and most of them adopt a modular architecture by which simple invertible modules are concatenated to attain a complex function.
This architecture is often called \textbf{Normalizing Flow (NF)}~\cite{papamakarios2019normalizing,tabak2013family,rippel2013high,dinh2014nice,dinh2016density}
and will be used in this work to implement the transform $f$.

\subsubsection{NDA model}

A key idea of NDA is to use an NF to map the distribution of one class in the $\pmb{x}$ space to the $\pmb{z}$ space,
where $p(\pmb{z})$ is a Gaussian.
After this transform, we can build a linear Gaussian model in the $\pmb{z}$ space to
describe the non-Gaussian observations in the $\pmb{x}$ space, as shown in Figure~\ref{fig:nda}.

As in PLDA, we assume that the prior $p(\pmb{\mu})$ for class mean $\pmb{\mu}$ is a diagonal Gaussian,
and the conditional distribution $p(\pmb{z}|\pmb{\mu})$ is a standard multivariate Gaussian whose covariance is set to $\mathbf{I}$.
This linear Gaussian model will represent the complex non-Gaussian data $\pmb{x}$ via the invertible transform ${f}$ represented by an NF.

It can be shown that the probability $p(\pmb{x}_1, ...,\pmb{x}_n)$ can be derived as follows:

\begin{eqnarray}
p(\pmb{x}_1, ..., \pmb{x}_n) &=& \int p(\pmb{x}_1, ..., \pmb{x}_n|\pmb{\mu}) p(\pmb{\mu}) \rm{d} \pmb{\mu} \nonumber \\
                             &=&  \prod_i^n J_{\pmb{x}_i} \int p(\pmb{z}_1, ..., \pmb{z}_n |\pmb{\mu}) p(\pmb{\mu}) \rm{d} \pmb{\mu} \nonumber\\
                             &=&  \prod_i^n J_{\pmb{x}_i} p(\pmb{z}_1, ..., \pmb{z}_n).
\end{eqnarray}
\noindent Note that the distribution $p(\pmb{\mu})$ and $p(\pmb{z}|\pmb{\mu})$ are both Gaussians,
and so $p(\pmb{z}_1, ..., \pmb{z}_n)$ can be easily computed as in PLDA, following Eq.~(\ref{eq:pxxx}).

This is a rather simple form and it indicates that if we can train an NF ${f}$,
the complex marginal distribution $p(\pmb{x}_1, ..., \pmb{x}_n)$ can be computed by transforming
the observation $\pmb{x}_i$ to a latent code $\pmb{z}_i$, and then computing the simple marginal
$p(\pmb{z}_1, ..., \pmb{z}_n)$ plus a correction term $\prod_i^n J_{\pmb{x}_i}$.
Interestingly, this correction term will be cancelled when computing the likelihood ratio for SRE scoring:

\begin{eqnarray}
R(\pmb{x}; \pmb{x}_1, ..., \pmb{x}_n) &=& \frac{p(\pmb{x},\pmb{x}_1, ..., \pmb{x}_n)}{p(\pmb{x}) p(\pmb{x}_1, ..., \pmb{x}_n)} \nonumber \\
                                      &=&  \frac{p(\pmb{z},\pmb{z}_1, ..., \pmb{z}_n)}{p(\pmb{z}) p(\pmb{z}_1, ..., \pmb{z}_n)}.  \nonumber
\end{eqnarray}

\noindent We therefore create a discriminant model which keeps the simple linear Gaussian form in the latent space,
but can deal with any complex within-class distributions. Finally, we note that if the transform $f$ is linear,
NDA falls back to PLDA.

\subsubsection{Model training}

The NDA model can be trained following the maximum-likelihood (ML) principle. Since all the speakers are
independent, the objective function is formulated by:

\begin{eqnarray}
\mathcal{L}({f}, \pmb{\epsilon}) &=& \sum_{k=1}^{K} \log p(\pmb{x}_1, ..., \pmb{x}_{n_k}) \nonumber \\
                                 &=& \sum_{k=1}^{K} \log \big\{\prod_i^{n_k} J_{\pmb{x}_i} p(\pmb{z}_1, ..., \pmb{z}_{n_k})\big\}, \nonumber
\end{eqnarray}

\noindent where $K$ is the number of speakers in the training data. During the training, firstly transform
the training samples $\pmb{x}_i$ to $\pmb{z}_i$, and then compute $J_{\pmb{x}_i}$ based on $f$
and $\pmb{x}_i$, secondly compute $p(\pmb{z}_1, ..., \pmb{z}_n)$ following Eq.~(\ref{eq:pxxx}).
Note that the covariance of $p(\pmb{z}|\pmb{\mu})$ has been fixed to $\mathbf{I}$ and so $\sigma$ is not a trainable parameter.

An important issue of the training algorithm is that for each speaker, all the data need to be
processed all at once. Therefore, the mini-batch design should be speaker-based. Moreover, we found
the training will be unstable if there are too few speakers in one mini-batch. We solve this problem by
postponing the model update when adequate speakers have been processed.

\section{Experiments}
\label{sec:exp}

\subsection{Data}

Three datasets were used in our experiments: VoxCeleb~\cite{nagrani2017voxceleb,chung2018voxceleb2},
SITW~\cite{mclaren2016speakers} and CNCeleb~\cite{fan2019cn}.
More information about these three datasets is presented below.

\emph{VoxCeleb}: This is a large-scale audiovisual speaker dataset collected by the University of Oxford, UK.
The entire database involves VoxCeleb1 and VoxCeleb2.
This dataset, after removing the utterances shared by SITW, was used to train the front-end x-vector, PLDA and NDA models.
The entire dataset contains $2,000$+ hours of speech signals from $7,000$+ speakers.
Data augmentation was applied to improve robustness,
with the MUSAN corpus~\cite{snyder2015musan} was used to generate noisy utterances,
and the room impulse responses (RIRS) corpus~\cite{ko2017study} was used to generate reverberant utterances.

\emph{SITW}: This is a standard evaluation dataset excerpted from VoxCeleb1, which consists of $299$ speakers.
In our experiments, both the \emph{Dev.Core} and \emph{Eval.Core} were used for evaluation.

\emph{CNCeleb}: This is a large-scale free speaker recognition dataset
collected by Tsinghua University.
It contains more than $130$k utterances from $1,000$ Chinese celebrities.
It covers $11$ diverse genres, which makes speaker recognition on this dataset much more challenging than on SITW.
The entire dataset was split into two parts: \emph{CNCeleb.Train}, which involves $111,257$ utterances from $800$ speakers,
was used to train the PLDA and the NDA models;
\emph{CNCeleb.Eval}, which involves $18,024$ utterances from $200$ speakers, was used for evaluation.

\subsection{Model Settings}

Our SRE system consists of two components: an x-vector frontend that produces speaker vectors,
and a scoring model that produces pair-wise scores to make genuine/imposter decisions.

\subsubsection{Front-end}


\textbf{x-vector}: The x-vector frontend was created using the Kaldi toolkit~\cite{povey2011kaldi}, following the VoxCeleb recipe.
The acoustic features are 40-dimensional Fbanks. The main architecture contains three components.
The first component is the feature-learning component, which involves $5$ time-delay (TD) layers to learn frame-level speaker features.
The slicing parameters for these $5$ TD layers are:
\{$t$-$2$, $t$-$1$, $t$, $t$+$1$, $t$+$2$\}, \{$t$-$2$, $t$, $t$+$2$\}, \{$t$-$3$, $t$, $t$+$3$\}, \{$t$\}, \{$t$\}.
The second component is the statistical pooling component,
which computes the mean and standard deviation of the frame-level features from a speech segment.
The final one is the speaker-classification component, which discriminates between different speakers.
This component has $2$ full-connection (FC) layers and the size of its output is $7,185$,
corresponding to the number of speakers in the training set.
Once trained, the $512$-dimensional activations of the penultimate FC layer are read out as an x-vector.



\subsubsection{Back-end}

\noindent \textbf{PLDA}: We implemented the standard PLDA model~\cite{Ioffe06} using the Kaldi toolkit~\cite{povey2011kaldi}.

\noindent \textbf{NDA}: We implemented the proposed NDA model in PyTorch.
The invertible transform $f$ was implemented using the \emph{RealNVP} architecture~\cite{dinh2016density}, a
particular NF that does not preserve the volume of the distribution.
We used $10$ non-volume preserving (NVP) layers, and the Adam optimizer~\cite{kingma2014adam} was used to train the model,
with the learning rate set to $0.001$.
For VoxCeleb, each mini-batch covers x-vectors from $600$ speakers, and for CNCeleb, each mini-batch
covers x-vectors from $200$ speakers.

\subsection{Basic results}

Experimental results on SITW Dev.Core, SITW Eval.Core and CNCeleb.Eval are shown in Table~\ref{tab:sitw}.
The results are reported in terms of equal error rate (EER) and minimum of the normalized detection cost function (minDCF) with two settings:
one with the prior target probability $P_{tar}$ set to $0.01$ (DCF($10^{-2}$)), and the other with $P_{tar}$
set to $0.001$ (DCF($10^{-3}$)).

Firstly, we focus on the full-dimensional PLDA (512) and NDA (512) scoring.
It can be observed that NDA scoring consistently outperformed PLDA scoring on the three evaluation datasets,
confirming that NDA is effective and more suitable as the x-vector back-end.
Besides, the performance of NDA on the CNCeleb.Eval is obviously better than that of PLDA (13.95\% vs. 12.51\%).
Considering the higher complexity of CNCeleb~\cite{fan2019cn}, this demonstrates that
NDA has better capability in dealing with complicated and challenging test conditions.

Secondly, we discarded some least discriminative dimensions, i.e., dimensions corresponding to the smallest $\epsilon_i$.
This approximates the subspace PLDA/NDA. The results are shown in Table~\ref{tab:sitw} as well.
It can be found that with this dimensionality reduction, performance improvement was generally observed.
Once again, NDA outperforms PLDA on almost all the datasets and with all the settings.

\begin{table}[htb!]
 \begin{center}
  \caption{Basic results on three evaluation datasets.}
   \label{tab:sitw}
    \scalebox{0.85}{
     \begin{tabular}{l|c|c|c|c|c}
       \hline
                          \multicolumn{6}{c}{SITW Dev.Core} \\
       \hline
             Front-end      &  Scoring    &    Dim         &  DCF($10^{-2}$)    &   DCF($10^{-3}$)   &   EER(\%) \\
       \hline
             x-vector       &  PLDA       &    512         &  0.485             &   0.704            &   4.082   \\
                            &  PLDA       &    300         &  0.380             &   0.581            &   3.389   \\
                       &  \textbf{PLDA}   & \textbf{150}   &  0.307             &   0.480            &   \textbf{3.196}   \\
                            &  NDA        &    512         &  0.480             &   0.720            &   4.043   \\
                            &  NDA        &    300         &  0.390             &   0.593            &   3.466   \\
                       &  \textbf{NDA}    & \textbf{150}   &  0.312             &   0.487            &   \textbf{3.196}   \\
       \hline
       \hline
                          \multicolumn{6}{c}{SITW Eval.Core} \\
       \hline
             Front-end      &  Scoring    &    Dim         &  DCF($10^{-2}$)    &   DCF($10^{-3}$)   &   EER(\%) \\
       \hline
             x-vector       &  PLDA       &    512         &  0.497             &   0.764            &   4.456   \\
                            &  PLDA       &    300         &  0.393             &   0.619            &   3.745   \\
                            &  PLDA       &    150         &  0.333             &   0.503            &   3.581   \\
                            &  NDA        &    512         &  0.494             &   0.771            &   4.155   \\
                            &  NDA        &    300         &  0.398             &   0.637            &   3.527   \\
                       &  \textbf{NDA}    & \textbf{150}   &  0.343             &   0.516            &   \textbf{3.417}   \\
       \hline
       \hline
                          \multicolumn{6}{c}{CNCeleb.Eval} \\
       \hline
             Front-end      &  Scoring    &    Dim         &  DCF($10^{-2}$)    &   DCF($10^{-3}$)   &   EER(\%) \\
       \hline
             x-vector       &  PLDA       &    512         &  0.691             &   0.837            &   13.95   \\
                            &  PLDA       &    300         &  0.674             &   0.822            &   13.72   \\
                            &  PLDA       &    150         &  0.660             &   0.816            &   13.63   \\
                            &  NDA        &    512         &  0.623             &   0.770            &   12.51   \\
                       &  \textbf{NDA}    & \textbf{300}   &  0.613             &   0.757            &   \textbf{12.45}   \\
                            &  NDA        &    150         &  0.612             &   0.752            &   12.60   \\
       \hline
     \end{tabular}}
 \end{center}
\end{table}

\subsection{Analysis for Gaussianality}

We have argued that the strength of NDA lies in the fact that the nonlinear transform $f$ can
map non-Gaussian observations $\pmb{x}$ to Gaussian latent codes $\pmb{z}$.
To test this argument, we compute the Gaussianality of the x-vectors before and after the NDA transform.
We compute the Skewness and Kurtosis for the marginal distribution (overall distribution without class labels),
conditional distribution (within-class distribution), and prior distribution (distribution of class means).
The results are reported in Table~\ref{tab:gauss}.




It can be seen that the values of Skewness and Kurtosis of the x-vectors
are substantially reduced after NDA transform, especially with the conditional distribution.
This is expected as the conditional distribution has been assumed to be Gaussian when NDA is designed and trained.
This improved Gaussianality allows a linear Gaussian model in the transformed space, as supposed by NDA.

\begin{table}[htb!]
 \begin{center}
  \caption{Gaussianality of x-vectors with/without NDA transform.}
   \label{tab:gauss}
    \scalebox{0.8}{
     \begin{tabular}{l|c|c|c|c|c|c|c}
       \hline
             \textbf{VoxCeleb}  &  & \multicolumn{2}{c|}{Marginal} & \multicolumn{2}{c|}{Conditional} & \multicolumn{2}{c}{Prior}  \\
       \hline
             Front-end &   NDA      &  Skew        &   Kurt      &  Skew        &   Kurt      &    Skew        &   Kurt       \\
       \hline
             x-vector  &   -        &  -0.087      &   -0.361    &  0.015       &   1.060     &   -0.045       &   -0.524     \\
                       &   +        &  0.015       &   0.134     &  \textbf{-0.004} & \textbf{0.267}  &   0.045        &   0.301      \\
       \hline
                      \multicolumn{8}{c}{}  \\
       \hline
             \textbf{CNCeleb}  &  & \multicolumn{2}{c|}{Marginal} & \multicolumn{2}{c|}{Conditional} & \multicolumn{2}{c}{Prior}  \\
       \hline
             Front-end &  NDA        &  Skew        &   Kurt        &  Skew        &   Kurt      &    Skew       &   Kurt        \\
       \hline
             x-vector  &  -          &  -0.139      &   0.180       &  -0.034      &   1.160     &   -0.160      &   -0.271      \\
                       &  +          &  0.002       &   0.122     &  \textbf{0.001}  &   \textbf{0.163}  &   -0.022      &   1.244       \\
       \hline
     \end{tabular}}
 \end{center}
\end{table}

\vspace{-1mm}

\subsection{Analysis for LDA pre-processing}

It is well known that LDA-based dimension reduction often provides significant
performance improvement for x-vector systems~\cite{li2019gaussian}.
Recently, the authors found that the contribution of LDA for x-vector systems
lies in normalization rather than discrimination. More specifically, for x-vectors,
the least discriminative dimensions coincide with the most non-Gaussian dimensions.
Therefore, LDA may improve the Gaussianality of x-vectors by discarding the least discriminative
dimensions~\cite{cai2020deep}.

Considering the success of the combination of LDA and PLDA, it is interesting to test
if LDA pre-processing contributes to NDA.
The results are shown in Table~\ref{tab:sitw},
where the dimensionality of the LDA projection space was set to $150$ for
VoxCeleb dataset and $300$ for CNCeleb dataset. These configurations delivered the
best performance with both PLDA and NDA.

It can be found that the performance was slightly improved after the LDA pre-processing,
with both PLDA and NDA. This is a bit surprising for NDA, as
NDA can deal with non-Gaussian data by itself, and so does not require an LDA to improve
the Gaussianality of the data. One possibility is that the reduced dimensionality
allows a better NDA modeling with limited data. However, more investigation is required.


\begin{table}[htb!]
 \begin{center}
  \caption{Performance with/without LDA pre-processing.}
   \label{tab:sitw}
    \scalebox{0.85}{
     \begin{tabular}{l|c|c|c|c}
       \hline
                          \multicolumn{5}{c}{SITW Dev.Core} \\
       \hline
             Front-end      &  Scoring                &  DCF($10^{-2}$)    &   DCF($10^{-3}$)   &   EER(\%) \\
       \hline
             x-vector       &  PLDA (150)             &  0.307             &   0.480            &   3.196   \\
                            &  NDA (150)              &  0.312             &   0.487            &   3.196   \\
       \hline
             x-vector       &  PLDA                   &  0.301             &   0.469            &   3.157   \\
             + LDA (150)    &  NDA                    &  0.295             &   0.472            &   \textbf{3.080}   \\
       \hline
       \hline
                          \multicolumn{5}{c}{SITW Eval.Core} \\
       \hline
             Front-end      &  Scoring                &  DCF($10^{-2}$)    &   DCF($10^{-3}$)   &   EER(\%) \\
       \hline
             x-vector       &  PLDA (150)             &  0.333             &   0.503            &   3.581   \\
                            &  NDA (150)              &  0.343             &   0.516            &   3.417   \\
       \hline
             x-vector       &  PLDA                   &  0.329             &   0.496            &   3.554   \\
             + LDA (150)    &  NDA                    &  0.335             &   0.508            &   \textbf{3.280}   \\
       \hline
       \hline
                          \multicolumn{5}{c}{CNCeleb.Eval} \\
       \hline
             Front-end      &  Scoring                &  DCF($10^{-2}$)    &   DCF($10^{-3}$)   &   EER(\%) \\
       \hline
             x-vector       &  PLDA (300)             &  0.674             &   0.822            &   13.72   \\
                            &  NDA (300)              &  0.613             &   0.757            &   12.45   \\
       \hline
             x-vector       &  PLDA                   &  0.675             &   0.821            &   13.73   \\
             + LDA (300)    &  NDA                    &  0.561             &   0.681            &   \textbf{12.28}   \\
       \hline
     \end{tabular}}
 \end{center}
\end{table}
\vspace{-1mm}

\section{Conclusions}
\label{sec:con}

We proposed a novel NDA model in this paper. It is a nonlinear extension of PLDA and can deal with
data with complex within-class distributions.
The key component of NDA is an NF-based invertible transform, which maps a complex distribution to a simple Gaussian
so that a linear Gaussian model can be established in the transformed space.
We applied NDA to SRE tasks and compared the performance with PLDA.
Results on the SITW and the CNCeleb datasets demonstrated that NDA can deliver
consistently better performance compared to PLDA.
Future work will investigate the joint training of the NDA scoring model and the speaker embedding model,
and apply NDA to raw acoustic features directly.



\newpage

\bibliographystyle{IEEEtran}
\bibliography{mybib}

\end{document}